%% file: TD-iDFT.tex
\documentclass[twocolumn,showpacs,amsmath,amssymb,floatfix,superscriptaddress,prl]{revtex4-1}
\usepackage{hyperref}
\usepackage[dvips]{graphicx}
\usepackage{amsfonts}
\usepackage{dcolumn}
\usepackage{bm}
\usepackage{color}
\usepackage{epsfig}
\usepackage{bbold}

\input{ourmacros}

\begin{document}

\title{AC transport in Correlated Quantum Dots: From Kondo to Coulomb 
blockade regime}

\author{G. Stefanucci}
\affiliation{Dipartimento di Fisica, Universit\`{a} di Roma Tor Vergata,
Via della Ricerca Scientifica 1, 00133 Rome, Italy; European Theoretical 
Spectroscopy Facility (ETSF)}
\affiliation{INFN, Laboratori Nazionali di Frascati, Via E. Fermi 40, 
00044 Frascati, Italy}

\author{S. Kurth}
\affiliation{Nano-Bio Spectroscopy Group and European Theoretical Spectroscopy 
Facility (ETSF), Dpto. de F\'{i}sica de Materiales,
Universidad del Pa\'{i}s Vasco UPV/EHU, Av. Tolosa 72, 
E-20018 San Sebasti\'{a}n, Spain}
\affiliation{IKERBASQUE, Basque Foundation for Science, Maria Diaz de Haro 3, 
E-48013 Bilbao, Spain}

\begin{abstract}
  We explore the finite bias DC differential conductance of a correlated
  quantum dot under the influence of an AC field, 
  from the low-temperature Kondo to the finite 
temperature Coulomb blockade regime. Real-time simulations 
are performed using a time-dependent generalization of the 
steady-state density functional theory (i-DFT) 
[Nano Lett. {\bf 15}, 8020 (2015)]. 
The numerical simplicity of 
i-DFT allows for unprecedented long time evolutions. Accurate values 
of average current and density are obtained by integrating over several 
periods of the AC field. We find that (i) the zero-temperature Kondo 
plateau is suppressed, (ii) the photon-assisted conductance peaks are 
shifted
due to correlations and  (iii)  the Coulomb blockade is 
lifted with a concomitant smoothening of the sharp diamond edges.
\end{abstract}
\maketitle


Over the last decades there have been tremendous  advances in
“manufacturing” nanoscale devices such as nano\-tubes, 
artificial atoms (quantum dots), or single molecules attached to metallic
leads \cite{me-book1,NazarovBlanter:09,me-book2,Baldea:16}.
The theoretical description of their transport properties often remains a
challenge since it involves the physics of strong correlations out of
thermal equilibrium. Even the (minimal) single impurity Anderson
model~\cite{Anderson:61} (SIAM), after almost sixty years
is still a  matter of intense research and debate. 
The development of clever analytical and numerical techniques has  
provided us with a fairly good understanding of its equilibrium and
steady state properties. Unfortunately, the same cannot be said for
time-dependent responses. 
In particular, very little is known about the behavior of the 
average current when an AC field is superimposed to a DC voltage. 
What is the fate of the zero-temperature Kondo plateau in the 
conductance? How does the charging energy affect the
photon-assisted conductance peaks? Is the finite-temperature Coulomb 
blockade lifted?
In this Letter we provide an answer to these and related fundamental 
questions.

One way to access AC (linear as well as nonlinear) transport 
properties 
consists in performing numerical simulations of the 
time-evolution of the system. Several time-propagation algorithms 
have been put forward. Time-dependent (TD) density matrix renormalization 
group~\cite{al2006adaptive,PhysRevB.78.195317,HeidrichMeisnerFeiguinDagotto:09,PhysRevB.88.045132,kirino2008time}, 
TD numerical renormalization group~\cite{PhysRevLett.95.196801},  
functional renormalization 
group~\cite{JakobsPletyukhovSchoeller:10,EckelHeidrichJakobsThorwartPletyokhovEgger:10,KarraschMedenSchoenhammer:10},
real-time Monte 
Carlo~\cite{PhysRevLett.100.176403,PhysRevLett.116.036801,WernerOkaEcksteinMillis:10,WernerOkaMillis:09,DroghettiRunger:16}, 
hierarchical equations of 
motion approach~\cite{cheng2015time}, iterative real-time path 
integral~\cite{PhysRevB.77.195316} and real-time effective 
action~\cite{bock2016buildup} have all been used to investigate 
the transient response to a sudden quench of bias or gate voltages,
and a reasonable agreement between them has been reached. 
These methods, however, are limited to short propagation times and 
AC responses are  difficult to address since convergence often 
requires averaging the TD current over several periods of the driving field. 
An alternative to time-propagation is the Floquet Green's function
approach.  SIAM responses to an oscillating {\em gate} have been reported in
Refs.~\cite{LopezAguadoPlateroTejedor:01,LopezAguadoPlateroTejedor:98} using
either a second-order self-energy (reasonable only at the particle-hole
symmetric point) or an analytically interpolated equilibrium self-energy 
(questionable for AC voltages). To the best of our knowledge, no 
attempts to calculate the SIAM conductance under AC {\em voltages} have 
been made so far.

Lately, strongly correlated systems have been studied also with density
functional theory (DFT) both in its static
\cite{SchoenhammerGunnarssonNoack:95,LimaSilvaOliveiraCapelle:03,MaletGoriGiorgi:12,LorenzanaYingBrosco:12,MirtschinkSeidlGoriGiorgi:13}
and time-dependent version
\cite{Verdozzi:08,kskvg.2010,UimonenKhosraviStanStefanucciKurthLeeuwenGross:11,KKPV.2013,HodgsonRamsdenChapmanLillystoneGodby:13,HodgsonRamsdenGodby:16}. 
In a recent work we proposed a steady-state density functional 
theory~\cite{StefanucciKurth:15} (i-DFT) to calculate the steady density on 
and the steady current through
a quantum junction sandwiched between metallic leads.
In Ref.~\cite{KurthStefanucci:16} we applied i-DFT to the SIAM and 
found excellent agreement with numerically exact methods 
in a wide range of bias and gate voltages, from weak to 
strong charging energies $U$ and from low to high temperatures $T$.

We here extend i-DFT to the time domain, thereby laying down a 
computationally feasible scheme to shed light on  
the AC transport properties of the SIAM. Our main findings are (i) 
AC voltage suppression of the $T=0$ Kondo plateau (ii) 
interaction-induced shift of the 
photon-assisted conductance peaks and 
(iii) lifting of the finite-temperature Coulomb blockade with 
concomitant smoothening of the diamond shape.

{\em Time-Dependent i-DFT - } 
i-DFT establishes a 
one-to-one correspondence between the pair density on and current 
through the impurity,
$(n,I)$, and the pair  
gate and bias voltages $(v,V)$. 
Accordingly, there exists a unique 
pair $(v_{s},V_{s})$ that in the noninteracting SIAM produces the 
same density and current as in the interacting one. The 
pair $(v_{s},V_{s})$ consists of the Kohn-Sham (KS) gate 
$v_{s}=v+v_{\rm H}+v_{\rm xc}$ and KS voltage $V_{s}=V+V_{\rm xc}$ where 
the Hartree potential $v_{\rm H}=Un$
whereas the exchange-correlation (xc) potentials $v_{\rm xc}$ and 
$V_{\rm xc}$ are universal functionals (i.e., independent of the 
external fields $v$ and $V$) of both density and current.
Knowledge of these functionals 
allows for calculating $(n,I)$ by solving self-consistently the equations
\be
n=2\sum_{\a=L,R}\int\frac{d\w}{2\p}f(\w-V_{s,\a})A_{\a}(\w)
\label{idft-n}
\ee
\be
I=2\int\frac{d\w}{2\p}\big[f(\w-V_{s,L})-f(\w-V_{s,R})\big]\callT(\w)
\label{idft-I}
\ee
where $V_{s,L}=-V_{s,R}=V_{s}/2$ is the bias applied to the left 
($L$) and right ($R$) leads and $f(\w)=1/(e^{\beta(\w-\m)}+1)$ is the 
Fermi function at inverse temperature $\b=1/T$ and chemical potential 
$\m$. 
In Eq.~(\ref{idft-n}) the partial spectral function  is given by
$A_{\a}(\w)\equiv G(\w)\G_{\a}(\w)G^{\dag}(\w)$ with the (retarded) KS Green's 
function $G(\w)=1/(\w-v-v_{\rm Hxc}-\S_{L}(\w)-\S_{R}(\w))$  and broadening 
$\G_{\a}(\w)=-2\Im\left[\S_{\a}(\w)\right]$, 
$\S_{\a}$ being the embedding self-energy of lead $\a$. The 
current in Eq.~(\ref{idft-I}) is expressed in terms of the KS 
transmission coefficient 
$\callT(\w)=A_{L}(\w)\G_{R}(\w)$.
The self-consistent solution of Eqs.~(\ref{idft-n}) and (\ref{idft-I})
is extremely efficient~\cite{KurthStefanucci:16,KurthStefanucci:17}.
It is therefore natural to extend i-DFT to the time domain and 
explore dynamical (as opposed to steady-state) 
transport properties. Setting the external bias $V(t)=\th(t)[V+V_{\rm 
AC}\sin(\W t)]$ we can provide a full characterization of the AC 
conductance.  

In the same spirit of the adiabatic approximations in standard TD-DFT
\cite{Ullrich:12,Maitra:16} we calculate the TD 
density $n(t)$ and current $I(t)$ by propagating the KS SIAM with 
TD gate potential $v_{s}(t)=v(t)+v_{\rm Hxc}[n(t),I(t)]$ and voltage 
$V_{s}(t)=V(t)+V_{\rm xc}[n(t),I(t)]$. 
The propagation is performed by solving the coupled TD KS equation
\be
i\frac{d}{dt}\left(\begin{array}{ccc}
\q_{L,k}\\ \q_{C,k} \\ \q_{R,k}
\end{array}\right)=
\left(\begin{array}{ccc}
h_{LL} & h_{LC} & 0\\ 
h_{CL} & h_{CC} & h_{CR}\\ 
0 & h_{RC} & h_{RR}
\end{array}\right)
\left(\begin{array}{ccc}
\q_{L,k}\\ \q_{C,k} \\ \q_{R,k}
\end{array}\right)
\label{TDKSeq}
\ee
for every KS state $\q_{k}$. Equation~(\ref{TDKSeq}) 
is written in a block form where the blocks $L$ and $R$ refer to the 
left and right leads whereas the block $C$ refers to the impurity 
site, hence $h_{CC}=v_{s}(t)$. We model the leads as 
semi-infinite tight-binding chains with nearest neighbor hopping $t_{\rm lead}$ 
and onsite energy $V_{s,\a}(t)$. Only the boundary site of the 
semi-infinite chains is connected to the impurity and the 
corresponding hopping 
amplitude is $t_{\rm link}$. We take the leads at 
half-filling ($\m=0$), choose both $t_{\rm lead}$ and 
$t_{\rm link}$ much larger than any other energy scale and set the ratio 
$t^{2}_{\rm link}/t_{\rm lead}=\g/2$ to stay in the 
wide band limit (WBL).

Equation~(\ref{TDKSeq}) is solved using  a generalization of the algorithm of 
Ref.~\onlinecite{ksarg.2005} to finite temperatures. The time propagation
is performed with a predictor-corrector scheme at each time 
step for $v_{\rm Hxc}$ and $V_{\rm xc}$. The TD occupation and 
current to plug into the xc potentials are obtained from the KS
wavefunctions according to
\be
n(t)=2\sum_{k}f(\e_{k})|\q_{C,k}(t)|^{2},
\ee
\be
I(t)=\frac{I_{L}(t)-I_{R}(t)}{2}.
\label{tdcurr}
\ee
The current at the $\a$ interface
\be
I_{\a}(t)=4\sum_{k}f(\e_{k})\Im\left[t_{\rm link}\q^{\ast}_{\a,k}(t)\q_{C,k}(t)\right]
\ee
is expressed in terms of the wavefunction $\q_{\a,k}(t)$ at
boundary site of lead $\a$. Notice that current conservation implies   
$I_{L}(t)+I_{R}(t)+\dot{n}(t)=0$ at all times.
For symmetric voltages $V_{L}(t)=-V_{R}(t)=V(t)/2$ and at the 
particle-hole symmetric point (ph--SP)  $v=-U/2$ we have $I_{L}(t)=-I_{R}(t)$.

In Ref.~\cite{KurthStefanucci:16} we proposed an accurate parametrization of 
$v_{\rm Hxc}$ and $V_{\rm xc}$. 
This parametrization has been shown to agree well with results from 
the functional and numerical renormalization 
group in a wide range of temperatures and 
interaction strengths. We have used the xc potentials
of Ref.~\cite{KurthStefanucci:16} to solve Eq.~(\ref{TDKSeq}).
A somewhat related current-dependent
approximation for $v_{\rm Hxc}$, valid for only one lead and at 
temperatures higher than the Kondo temperature,
has recently been suggested in
Ref.~\onlinecite{DittmannSplettstoesserHelbig}.

{\em Results - } 
We consider the SIAM initially (times $t\leq0$) in thermal equilibrium
and then driven out of equilibrium by a symmetric bias $V_{L}(t) 
= - V_{R}(t) = V(t)/2$. The bias  is 
the sum of a DC and an AC contribution, i.e.,  
\be
V(t) = V + V_{\rm AC} \sin(\W t)  \mbox{\hspace*{1cm} for $t>0$.}
\label{td_bias}
\ee
After a sufficiently long propagation time $t_{\ell}$
the transient features die  out 
and all physical observables become periodic functions of time.
We then calculate the DC component of the current by averaging
$I(t)$ over $N$ periods $\tau=2 \pi/\W$, i.e.,
$I_{\rm DC}=\int_{t_{\ell}}^{t_{\ell}+N\t}I(t)/(N\t)$. 
Depending on the parameters, a good convergence may require $N$ of 
the order of $10$ or larger.
Finally we calculate the finite-bias DC conductance 
$G={\rm d}I_{\rm DC}/{\rm d} V$, highlighting its most relevant 
features as $V_{\rm AC}$ and $\W$ are varied.
We consider over two physically distinct regimes: the Kondo regime
at zero temperature and the Coulomb blockade (CB) regime at
temperatures larger than the Kondo temperature. 

First we checked that for $U=V=0$ our TD algorithm 
agrees with the exact (in the WBL approximation) noninteracting formula 
\cite{JauhoWingreenMeir:94}
\be
G^{U=0} = \sum_{k=-\infty}^{\infty} J_k^2\left(-\frac{V_{\rm AC}}{\W} \right)
\frac{\frac{\gamma^2}{4}}{(v - k \W)^2 + \frac{\gamma^2}{4}}
\label{conduct_nonint}
\ee
where $J_k(x)$ is the Bessel function of order $k$. 
Without interactions  photon-assisted
transport (PAT), i.e., charge transfer accompanied
by the emission or absorption of photons, is the only scattering 
mechanism.

\begin{figure}[t]
\includegraphics[width=0.47\textwidth]{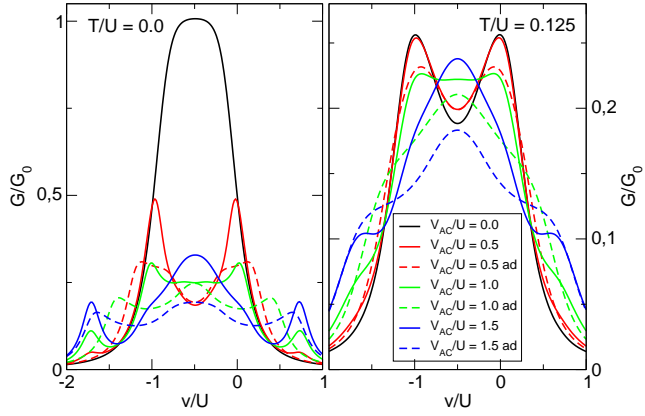}
\caption{Zero-bias DC conductance (solid lines) as well as its adiabatic
  counterpart of Eq.~(\protect\ref{diff_cond_ad}) (dashed) versus $v$
  for different AC amplitudes. The SIAM parameters are  $U/\gamma=4$, 
  $\W/U=\pi/4$ and temperature $T=0$ (left panel) and $T/U=0.125$ 
  (right panel). $G_0=\frac{1}{\pi}$ is the quantum of conductance.}
\label{dc_cond_ac}
\end{figure}

Turning on the interaction the scenario changes dramatically,
especially for gates $v$ in the range $(-U,0)$ where correlation 
effects are enhanced.
In Fig.~\ref{dc_cond_ac} we show results 
for the zero-bias DC conductance
as function of $v$ for different  
$V_{\rm AC}$. At temperature $T=0$
(left panel) and already for small 
$V_{\rm AC}$ the Kondo plateau is drastically suppressed. 
The DC conductance 
resembles the one at $V_{\rm AC}=0$ in the CB 
regime with two peaks separated by roughly the charging energy $U$ 
and a minimum at the ph--SP.
The suppression of Kondo correlations has already 
been observed in Refs.~\cite{GoldinAvishai:98,GoldinAvishai:00} using 
a perturbative approach as well as in
Refs.~ \cite{LopezAguadoPlateroTejedor:98,LopezAguadoPlateroTejedor:01}
where, rather than an AC bias, the {\em gate} 
potential is perturbed harmonically.
As $V_{\rm AC}$ increases, however, the minimum is 
first converted into a plateau and eventually into a maximum with the 
concomitant  washing out  of the side peaks.
The decrease and subsequent increase of the Kondo peak
at the ph--SP predicted by our calculations 
is in agreement with experimental findings 
reported in Ref.~\cite{Elzerman:00}.

In the CB regime ($T/U=0.125$, right panel) a small 
AC bias barely changes the values of $G$ at $V_{\rm AC}=0$.
However, for $V_{\rm AC}\approx U$  the DC conductance 
develops a plateau  and a 
maximum at the ph--SP for even larger amplitudes. 
In both regimes, see left and right panels, 
we also observe increasingly higher side peaks at
$v\approx \W$ and $v\approx -U-\W$ with increasing $V_{\rm AC}$, a 
clear signature of PAT.
We will discuss the precise position of these peaks below.

To highlight nonadiabatic (memory) effects
we define the ``adiabatic'' DC conductance according to
\be
G_{\rm ad} = \frac{1}{\tau} \int_0^{\tau} {\rm d} t \; G_{\rm 
DC}(V(t)) ,
\label{diff_cond_ad}
\ee
where $G_{\rm DC}(V)=\left.{\rm d}I/{\rm d}V\right|_{V_{\rm AC}=0}$
is the finite-bias DC conductance at vanishing AC amplitude.
The results for $G_{\rm ad}$ are
shown in both panels of Fig.~\ref{dc_cond_ac} (dashed lines). For small 
$V_{\rm AC}$ and around the ph--SP,
$G_{\rm ad}$ is remarkably close to $G$ both in the Kondo and the CB regime
while away from ph--SP the two quantities only share qualitative features
at best. For larger $V_{\rm AC}$, adiabatic and non-adiabatic
DC conductances increasingly differ, pointing to the importance of memory 
effects.

\begin{figure}[t]
\includegraphics[width=0.47\textwidth]{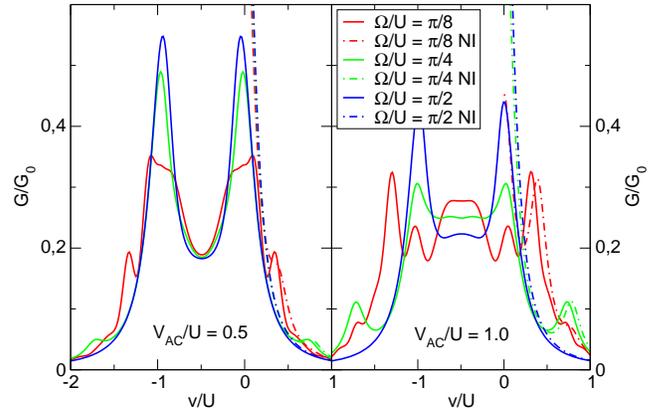}
\caption{Zero-bias DC conductance versus $v$ for different AC 
frequencies and for amplitudes $V_{\rm AC}/U=0.5$ (left panel) and 
$V_{\rm AC}/U=1.0$ (right panel). Other parameters as in Fig.~\ref{dc_cond_ac}.}
\label{dc_cond_ac_om}
\end{figure}

\begin{figure}[tb]
\includegraphics[width=0.47\textwidth]{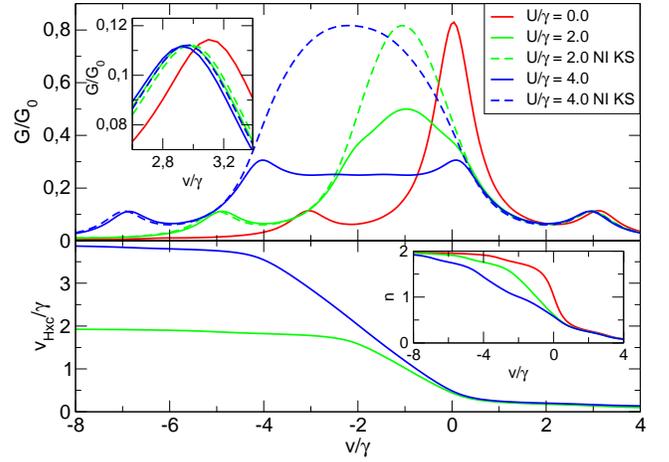}
\caption{Upper panel: Zero-bias DC conductances (solid) versus $v$ for
$\W/\g=\pi$, $V_{\rm AC}/\g=4.0$ and 
different values of $U$. Dashed lines are the noninteracting
result of Eq.~(\protect\ref{conduct_nonint}) with $v$ 
replaced by the time-averaged KS potential $v_{s}$. Inset:
zoom in around $v=\W$ showing shifts in the peak
position of the PAT peak. Lower panel: time-averaged Hxc gate
and density (inset).}
\label{diff_U}
\end{figure}

\begin{figure*}[t]
\includegraphics[width=0.98\textwidth]{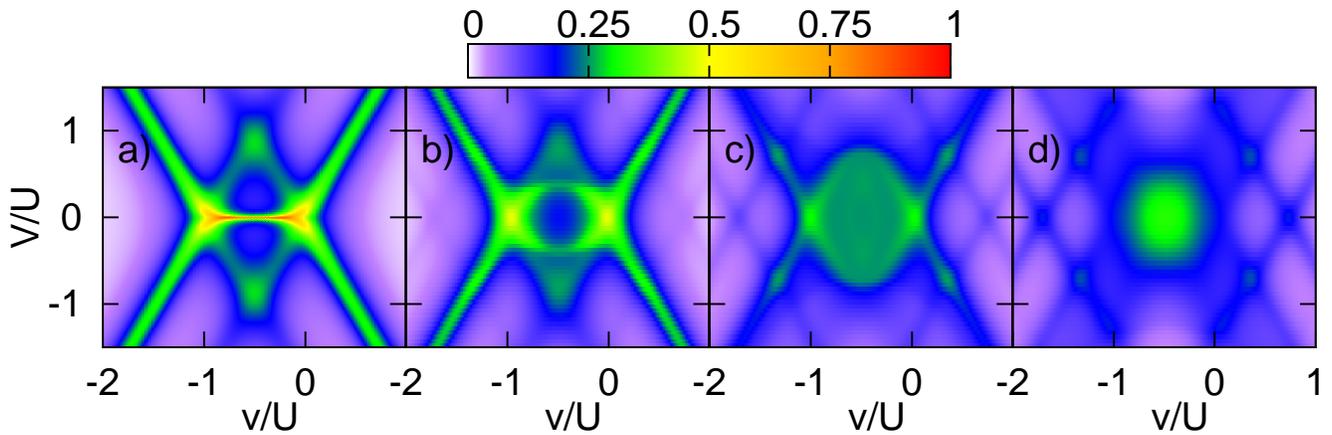}
\caption{DC conductance in the $(v,V)$ plane at
zero temperature  for a) $V_{\rm AC}/U=0.0$, b) $V_{\rm AC}/U=0.5$, c) 
$V_{\rm AC}/U=1.0$ and d) $V_{\rm AC}/U=1.5$. Other parameters as in 
Fig.~\ref{dc_cond_ac}.}
\label{didv_ac_siam}
\end{figure*}

In Fig.~\ref{dc_cond_ac_om} we show the dependence of the zero-bias $G$ at
$T=0$ on the AC frequency for two different AC amplitudes. 
At the small amplitude $V_{\rm AC}/U=0.5$ (left panel) 
the behavior of $G$ is qualitatively similar for different 
frequencies (suppression of the 
Kondo plateau and arising of PAT peaks).
Interestingly, for the larger amplitude 
$V_{\rm AC}/U=1.0$ (right panel), $G$ 
develops a plateau around the ph--SP {\em for all} 
frequencies, with a height that  decreases monotonically with $\W$.
For this larger amplitude the PAT peaks 
emerge more distinctly and we clearly appreciate the nonlinear (in 
$V_{\rm AC}$) effect of
absorption/emission of two photons (for $\W/U=\p/8$).

The dashed curves starting at $v=0$ in both panels of Fig.~\ref{dc_cond_ac_om} 
are the values of $G^{U=0}$, see Eq.~(\ref{conduct_nonint})
(which is an even functions of $v$),
exhibiting PAT peaks at $v=k \W$ with $k$ integer~\cite{JauhoWingreenMeir:94}.
For $v>0$, the interacting and noninteracing results agree reasonably well
while for $v<0$ correlations shift the PAT peaks by approximately 
$-U$.

It is worth noting that 
in the interacting case the PAT peak at $v\approx \W$ is
slightly shifted  to lower gates. To shed some light on this shift
we calculated the zero-bias $G$ for various charging energies $U$. 
The results are shown in the upper panel of Fig.~\ref{diff_U}
where the inset is a magnification of the PAT peak at $v\approx \W$. 
For these gate values correlations are weak as confirmed by the small 
value $n\approx0.15$ of the time-averaged density, see inset in the 
lower panel.
In the weakly correlated regime the DC conductance is well 
approximated by the noninteracting formula of 
Eq.~(\ref{conduct_nonint}). However, from the i-DFT perspective, the $v$ 
in Eq.~(\ref{conduct_nonint}) is not just the bare gate but instead the time-averaged 
KS gate $v_s=v+v_{\rm Hxc}$, i.e., the sum of the bare gate and the time-averaged 
Hxc potential $v_{\rm Hxc}$ shown in the lower panel of 
Fig.~\ref{diff_U}. Consequently, the PAT peak in $G$ 
occurs at $v=\W-v_{\rm Hxc}$, i.e., it is
shifted to lower energies. This is precisely the shift we observe
in our calculations. 
In fact, if we evaluate Eq.~(\ref{conduct_nonint}) making the replacement
$v\to v_s$ we get a remarkable good agreement with the 
{\em interacting} $G$ (including position and height of the 
PAT peaks) for $v$ outside the interval $(-U,0)$, see dashed curves in upper panel of 
Fig.~\ref{diff_U}.
Obviously, this good agreement breaks
down for gates in the interval $(-U,0)$ since the 
system can no longer
be considered weakly correlated. 

So far we presented results for the zero-bias DC conductance.
However, the proposed method is not limited to this very special case. 
In Fig.~\ref{didv_ac_siam} we display the finite-bias $G$ as
function of $v$ and $V$ for different amplitudes of the 
AC field. The calculations 
are done at temperature $T=0$. To appreciate the effects of the 
AC field the well-known pure DC result, i.e., $V_{\rm AC}=0$, is 
shown in panel a). We correctly reproduce the CB diamond 
as well as the Kondo resonance at $V=0$. 
A moderate amplitude of the AC field, panel b), leads to a 
suppression of the Kondo resonance (see also Fig.~\ref{dc_cond_ac}). 
Moreover, starting from the finite bias corners  
at $v/U=-1/2$ and $V/U=\pm1$, the CB is lifted inside the diamond.
As $V_{\rm AC}$ increases further, panels c) and d), 
the CB is lifted everywhere in the diamond, leaving for the largest 
amplitude, panel d), an area of increased conductance around 
the ph--SP at $V=0$. We also find that the side peaks due to PAT
seen in Figs.~\ref{dc_cond_ac} and \ref{dc_cond_ac_om} 
branch at finite bias into straight lines with negative and positive slopes, 
reminiscent of the typical $G$ lines at $V_{\rm AC}=0$. Not surprisingly, 
these lines become more pronounced as $V_{\rm AC}$ increases. 

In summary, we have investigated electron transport through a correlated
quantum dot subjected simultaneously to DC and AC biases
by explicit and numerically efficient time propagation in a DFT framework. 
For the corresponding Hxc gate and xc bias potentials we have suggested a
time-local approximation based on recently proposed accurate
i-DFT functionals for the steady state. We find that in the Kondo regime 
already for small AC amplitude the Kondo plateau in the zero-bias
DC conductance is strongly suppressed while in the CB regime the changes
are less pronounced. The observed shift of the photon-assisted transport
peaks due to electronic interactions can readily be explained within DFT.
At finite DC bias, Coulomb blockade is lifted with increasing
AC amplitude and the CB diamond is deformed.

G.S. acknowledges funding by MIUR FIRB Grant No. RBFR12SW0J, EC funding
through the RISE Co-ExAN (GA644076), and funding through the INFN-Nemesys
project.
S.K. acknowledges funding by a grant of the "Ministerio de Economia y
Competividad (MINECO)" (FIS2016-79464-P) and by the
``Grupos Consolidados UPV/EHU del Gobierno Vasco'' (IT578-13).

\end{document}

%% file: ourmacros.tex
\newcommand{\be}{\begin{equation}}
\newcommand{\ee}{\end{equation}}
\newcommand{\bea}{\begin{eqnarray}}
\newcommand{\eea}{\end{eqnarray}}

\def\a{\alpha}
\def\b{\beta}
\def\g{\gamma}
\def\G{\Gamma}

\def\e{\epsilon}

\def\th{\theta}

\def\m{\mu}

\def\p{\pi}

\def\S{\Sigma}
\def\t{\tau}

\def\w{\omega}
\def\W{\Omega}
\def\q{\psi}

\def\callT{\mbox{$\mathcal{T}$}}

\def\Im{{\rm Im}}

\def\1op{\hat{\mathbbm{1}}}

%% file: TD-iDFT.bbl
\begin{thebibliography}{46}%
\makeatletter
\providecommand \@ifxundefined [1]{%
 \@ifx{#1\undefined}
}%
\providecommand \@ifnum [1]{%
 \ifnum #1\expandafter \@firstoftwo
 \else \expandafter \@secondoftwo
 \fi
}%
\providecommand \@ifx [1]{%
 \ifx #1\expandafter \@firstoftwo
 \else \expandafter \@secondoftwo
 \fi
}%
\providecommand \natexlab [1]{#1}%
\providecommand \enquote  [1]{``#1''}%
\providecommand \bibnamefont  [1]{#1}%
\providecommand \bibfnamefont [1]{#1}%
\providecommand \citenamefont [1]{#1}%
\providecommand \href@noop [0]{\@secondoftwo}%
\providecommand \href [0]{\begingroup \@sanitize@url \@href}%
\providecommand \@href[1]{\@@startlink{#1}\@@href}%
\providecommand \@@href[1]{\endgroup#1\@@endlink}%
\providecommand \@sanitize@url [0]{\catcode `\\12\catcode `\$12\catcode
  `\&12\catcode `\#12\catcode `\^12\catcode `\_12\catcode `\%12\relax}%
\providecommand \@@startlink[1]{}%
\providecommand \@@endlink[0]{}%
\providecommand \url  [0]{\begingroup\@sanitize@url \@url }%
\providecommand \@url [1]{\endgroup\@href {#1}{\urlprefix }}%
\providecommand \urlprefix  [0]{URL }%
\providecommand \Eprint [0]{\href }%
\providecommand \doibase [0]{http://dx.doi.org/}%
\providecommand \selectlanguage [0]{\@gobble}%
\providecommand \bibinfo  [0]{\@secondoftwo}%
\providecommand \bibfield  [0]{\@secondoftwo}%
\providecommand \translation [1]{[#1]}%
\providecommand \BibitemOpen [0]{}%
\providecommand \bibitemStop [0]{}%
\providecommand \bibitemNoStop [0]{.\EOS\space}%
\providecommand \EOS [0]{\spacefactor3000\relax}%
\providecommand \BibitemShut  [1]{\csname bibitem#1\endcsname}%
\let\auto@bib@innerbib\@empty
\bibitem [{\citenamefont {Cuniberti}\ \emph {et~al.}(2005)\citenamefont
  {Cuniberti}, \citenamefont {Fagas},\ and\ \citenamefont
  {Richter}}]{me-book1}%
  \BibitemOpen
  \bibfield  {author} {\bibinfo {author} {\bibfnamefont {G.}~\bibnamefont
  {Cuniberti}}, \bibinfo {author} {\bibfnamefont {G.}~\bibnamefont {Fagas}}, \
  and\ \bibinfo {author} {\bibfnamefont {K.}~\bibnamefont {Richter}},\
  }\href@noop {} {\emph {\bibinfo {title} {Introducing Molecular
  Electronics}}}\ (\bibinfo  {publisher} {Springer},\ \bibinfo {address}
  {Heidelberg},\ \bibinfo {year} {2005})\BibitemShut {NoStop}%
\bibitem [{\citenamefont {Nazarov}\ and\ \citenamefont
  {Blanter}(2009)}]{NazarovBlanter:09}%
  \BibitemOpen
  \bibfield  {author} {\bibinfo {author} {\bibfnamefont {Y.~V.}\ \bibnamefont
  {Nazarov}}\ and\ \bibinfo {author} {\bibfnamefont {Y.~M.}\ \bibnamefont
  {Blanter}},\ }\href@noop {} {\emph {\bibinfo {title} {Quantum Transport:
  Introduction to Nanoscience}}}\ (\bibinfo  {publisher} {Cambridge University
  Press},\ \bibinfo {address} {Cambridge},\ \bibinfo {year} {2009})\BibitemShut
  {NoStop}%
\bibitem [{\citenamefont {Cuevas}\ and\ \citenamefont
  {Scheer}(2010)}]{me-book2}%
  \BibitemOpen
  \bibfield  {author} {\bibinfo {author} {\bibfnamefont {J.~C.}\ \bibnamefont
  {Cuevas}}\ and\ \bibinfo {author} {\bibfnamefont {E.}~\bibnamefont
  {Scheer}},\ }\href@noop {} {\emph {\bibinfo {title} {Molecular Electronics:
  An Introduction to Theory and Experiment}}}\ (\bibinfo  {publisher} {World
  Scientific},\ \bibinfo {address} {London},\ \bibinfo {year}
  {2010})\BibitemShut {NoStop}%
\bibitem [{\citenamefont {Baldea}(2015)}]{Baldea:16}%
  \BibitemOpen
  \bibfield  {author} {\bibinfo {author} {\bibfnamefont {I.}~\bibnamefont
  {Baldea}},\ }\href@noop {} {\emph {\bibinfo {title} {Molecular Electronics:
  An Experimental and Theoretical Approach}}}\ (\bibinfo  {publisher} {Pan
  Stanford},\ \bibinfo {year} {2015})\BibitemShut {NoStop}%
\bibitem [{\citenamefont {Anderson}(1961)}]{Anderson:61}%
  \BibitemOpen
  \bibfield  {author} {\bibinfo {author} {\bibfnamefont {P.~W.}\ \bibnamefont
  {Anderson}},\ }\href {\doibase 10.1103/PhysRev.124.41} {\bibfield  {journal}
  {\bibinfo  {journal} {Phys. Rev.}\ }\textbf {\bibinfo {volume} {124}},\
  \bibinfo {pages} {41} (\bibinfo {year} {1961})}\BibitemShut {NoStop}%
\bibitem [{\citenamefont {Al-Hassanieh}\ \emph {et~al.}(2006)\citenamefont
  {Al-Hassanieh}, \citenamefont {Feiguin}, \citenamefont {Riera}, \citenamefont
  {B{\"u}sser},\ and\ \citenamefont {Dagotto}}]{al2006adaptive}%
  \BibitemOpen
  \bibfield  {author} {\bibinfo {author} {\bibfnamefont {K.~A.}\ \bibnamefont
  {Al-Hassanieh}}, \bibinfo {author} {\bibfnamefont {A.~E.}\ \bibnamefont
  {Feiguin}}, \bibinfo {author} {\bibfnamefont {J.~A.}\ \bibnamefont {Riera}},
  \bibinfo {author} {\bibfnamefont {C.}~\bibnamefont {B{\"u}sser}}, \ and\
  \bibinfo {author} {\bibfnamefont {E.}~\bibnamefont {Dagotto}},\ }\href
  {https://link.aps.org/doi/10.1103/PhysRevB.73.195304} {\bibfield  {journal}
  {\bibinfo  {journal} {Phys. Rev. B}\ }\textbf {\bibinfo {volume} {73}},\
  \bibinfo {pages} {195304} (\bibinfo {year} {2006})}\BibitemShut {NoStop}%
\bibitem [{\citenamefont {Dias~da Silva}\ \emph {et~al.}(2008)\citenamefont
  {Dias~da Silva}, \citenamefont {Heidrich-Meisner}, \citenamefont {Feiguin},
  \citenamefont {B\"usser}, \citenamefont {Martins}, \citenamefont {Anda},\
  and\ \citenamefont {Dagotto}}]{PhysRevB.78.195317}%
  \BibitemOpen
  \bibfield  {author} {\bibinfo {author} {\bibfnamefont {L.~G. G.~V.}\
  \bibnamefont {Dias~da Silva}}, \bibinfo {author} {\bibfnamefont
  {F.}~\bibnamefont {Heidrich-Meisner}}, \bibinfo {author} {\bibfnamefont
  {A.~E.}\ \bibnamefont {Feiguin}}, \bibinfo {author} {\bibfnamefont {C.~A.}\
  \bibnamefont {B\"usser}}, \bibinfo {author} {\bibfnamefont {G.~B.}\
  \bibnamefont {Martins}}, \bibinfo {author} {\bibfnamefont {E.~V.}\
  \bibnamefont {Anda}}, \ and\ \bibinfo {author} {\bibfnamefont
  {E.}~\bibnamefont {Dagotto}},\ }\href {\doibase 10.1103/PhysRevB.78.195317}
  {\bibfield  {journal} {\bibinfo  {journal} {Phys. Rev. B}\ }\textbf {\bibinfo
  {volume} {78}},\ \bibinfo {pages} {195317} (\bibinfo {year}
  {2008})}\BibitemShut {NoStop}%
\bibitem [{\citenamefont {Heidrich-Meisner}\ \emph {et~al.}(2009)\citenamefont
  {Heidrich-Meisner}, \citenamefont {Feiguin},\ and\ \citenamefont
  {Dagotto}}]{HeidrichMeisnerFeiguinDagotto:09}%
  \BibitemOpen
  \bibfield  {author} {\bibinfo {author} {\bibfnamefont {F.}~\bibnamefont
  {Heidrich-Meisner}}, \bibinfo {author} {\bibfnamefont {A.~E.}\ \bibnamefont
  {Feiguin}}, \ and\ \bibinfo {author} {\bibfnamefont {E.}~\bibnamefont
  {Dagotto}},\ }\href {\doibase 10.1103/PhysRevB.79.235336} {\bibfield
  {journal} {\bibinfo  {journal} {Phys. Rev. B}\ }\textbf {\bibinfo {volume}
  {79}},\ \bibinfo {pages} {235336} (\bibinfo {year} {2009})}\BibitemShut
  {NoStop}%
\bibitem [{\citenamefont {Nuss}\ \emph {et~al.}(2013)\citenamefont {Nuss},
  \citenamefont {Ganahl}, \citenamefont {Evertz}, \citenamefont {Arrigoni},\
  and\ \citenamefont {von~der Linden}}]{PhysRevB.88.045132}%
  \BibitemOpen
  \bibfield  {author} {\bibinfo {author} {\bibfnamefont {M.}~\bibnamefont
  {Nuss}}, \bibinfo {author} {\bibfnamefont {M.}~\bibnamefont {Ganahl}},
  \bibinfo {author} {\bibfnamefont {H.~G.}\ \bibnamefont {Evertz}}, \bibinfo
  {author} {\bibfnamefont {E.}~\bibnamefont {Arrigoni}}, \ and\ \bibinfo
  {author} {\bibfnamefont {W.}~\bibnamefont {von~der Linden}},\ }\href
  {\doibase 10.1103/PhysRevB.88.045132} {\bibfield  {journal} {\bibinfo
  {journal} {Phys. Rev. B}\ }\textbf {\bibinfo {volume} {88}},\ \bibinfo
  {pages} {045132} (\bibinfo {year} {2013})}\BibitemShut {NoStop}%
\bibitem [{\citenamefont {Kirino}\ \emph {et~al.}(2008)\citenamefont {Kirino},
  \citenamefont {Fujii}, \citenamefont {Zhao},\ and\ \citenamefont
  {Ueda}}]{kirino2008time}%
  \BibitemOpen
  \bibfield  {author} {\bibinfo {author} {\bibfnamefont {S.}~\bibnamefont
  {Kirino}}, \bibinfo {author} {\bibfnamefont {T.}~\bibnamefont {Fujii}},
  \bibinfo {author} {\bibfnamefont {J.}~\bibnamefont {Zhao}}, \ and\ \bibinfo
  {author} {\bibfnamefont {K.}~\bibnamefont {Ueda}},\ }\href
  {https://doi.org/10.1143/JPSJ.77.084704} {\bibfield  {journal} {\bibinfo
  {journal} {J. Phys. Soc. Jpn.}\ }\textbf {\bibinfo {volume} {77}},\ \bibinfo
  {pages} {084704} (\bibinfo {year} {2008})}\BibitemShut {NoStop}%
\bibitem [{\citenamefont {Anders}\ and\ \citenamefont
  {Schiller}(2005)}]{PhysRevLett.95.196801}%
  \BibitemOpen
  \bibfield  {author} {\bibinfo {author} {\bibfnamefont {F.~B.}\ \bibnamefont
  {Anders}}\ and\ \bibinfo {author} {\bibfnamefont {A.}~\bibnamefont
  {Schiller}},\ }\href {\doibase 10.1103/PhysRevLett.95.196801} {\bibfield
  {journal} {\bibinfo  {journal} {Phys. Rev. Lett.}\ }\textbf {\bibinfo
  {volume} {95}},\ \bibinfo {pages} {196801} (\bibinfo {year}
  {2005})}\BibitemShut {NoStop}%
\bibitem [{\citenamefont {Jakobs}\ \emph {et~al.}(2010)\citenamefont {Jakobs},
  \citenamefont {Pletyukhov},\ and\ \citenamefont
  {Schoeller}}]{JakobsPletyukhovSchoeller:10}%
  \BibitemOpen
  \bibfield  {author} {\bibinfo {author} {\bibfnamefont {S.~G.}\ \bibnamefont
  {Jakobs}}, \bibinfo {author} {\bibfnamefont {M.}~\bibnamefont {Pletyukhov}},
  \ and\ \bibinfo {author} {\bibfnamefont {H.}~\bibnamefont {Schoeller}},\
  }\href {\doibase 10.1103/PhysRevB.81.195109} {\bibfield  {journal} {\bibinfo
  {journal} {Phys. Rev. B}\ }\textbf {\bibinfo {volume} {81}},\ \bibinfo
  {pages} {195109} (\bibinfo {year} {2010})}\BibitemShut {NoStop}%
\bibitem [{\citenamefont {Eckel}\ \emph {et~al.}(2010)\citenamefont {Eckel},
  \citenamefont {Heidrich-Meisner}, \citenamefont {Jakobs}, \citenamefont
  {Thorwart}, \citenamefont {Pletyukhov},\ and\ \citenamefont
  {Egger}}]{EckelHeidrichJakobsThorwartPletyokhovEgger:10}%
  \BibitemOpen
  \bibfield  {author} {\bibinfo {author} {\bibfnamefont {J.}~\bibnamefont
  {Eckel}}, \bibinfo {author} {\bibfnamefont {F.}~\bibnamefont
  {Heidrich-Meisner}}, \bibinfo {author} {\bibfnamefont {S.~G.}\ \bibnamefont
  {Jakobs}}, \bibinfo {author} {\bibfnamefont {M.}~\bibnamefont {Thorwart}},
  \bibinfo {author} {\bibfnamefont {M.}~\bibnamefont {Pletyukhov}}, \ and\
  \bibinfo {author} {\bibfnamefont {R.}~\bibnamefont {Egger}},\ }\href
  {http://stacks.iop.org/1367-2630/12/i=4/a=043042} {\bibfield  {journal}
  {\bibinfo  {journal} {New J. Phys.}\ }\textbf {\bibinfo {volume} {12}},\
  \bibinfo {pages} {043042} (\bibinfo {year} {2010})}\BibitemShut {NoStop}%
\bibitem [{\citenamefont {Karrasch}\ \emph {et~al.}(2010)\citenamefont
  {Karrasch}, \citenamefont {Meden},\ and\ \citenamefont
  {Sch\"onhammer}}]{KarraschMedenSchoenhammer:10}%
  \BibitemOpen
  \bibfield  {author} {\bibinfo {author} {\bibfnamefont {C.}~\bibnamefont
  {Karrasch}}, \bibinfo {author} {\bibfnamefont {V.}~\bibnamefont {Meden}}, \
  and\ \bibinfo {author} {\bibfnamefont {K.}~\bibnamefont {Sch\"onhammer}},\
  }\href {\doibase 10.1103/PhysRevB.82.125114} {\bibfield  {journal} {\bibinfo
  {journal} {Phys. Rev. B}\ }\textbf {\bibinfo {volume} {82}},\ \bibinfo
  {pages} {125114} (\bibinfo {year} {2010})}\BibitemShut {NoStop}%
\bibitem [{\citenamefont {M\"uhlbacher}\ and\ \citenamefont
  {Rabani}(2008)}]{PhysRevLett.100.176403}%
  \BibitemOpen
  \bibfield  {author} {\bibinfo {author} {\bibfnamefont {L.}~\bibnamefont
  {M\"uhlbacher}}\ and\ \bibinfo {author} {\bibfnamefont {E.}~\bibnamefont
  {Rabani}},\ }\href {\doibase 10.1103/PhysRevLett.100.176403} {\bibfield
  {journal} {\bibinfo  {journal} {Phys. Rev. Lett.}\ }\textbf {\bibinfo
  {volume} {100}},\ \bibinfo {pages} {176403} (\bibinfo {year}
  {2008})}\BibitemShut {NoStop}%
\bibitem [{\citenamefont {Antipov}\ \emph {et~al.}(2016)\citenamefont
  {Antipov}, \citenamefont {Dong},\ and\ \citenamefont
  {Gull}}]{PhysRevLett.116.036801}%
  \BibitemOpen
  \bibfield  {author} {\bibinfo {author} {\bibfnamefont {A.~E.}\ \bibnamefont
  {Antipov}}, \bibinfo {author} {\bibfnamefont {Q.}~\bibnamefont {Dong}}, \
  and\ \bibinfo {author} {\bibfnamefont {E.}~\bibnamefont {Gull}},\ }\href
  {\doibase 10.1103/PhysRevLett.116.036801} {\bibfield  {journal} {\bibinfo
  {journal} {Phys. Rev. Lett.}\ }\textbf {\bibinfo {volume} {116}},\ \bibinfo
  {pages} {036801} (\bibinfo {year} {2016})}\BibitemShut {NoStop}%
\bibitem [{\citenamefont {Werner}\ \emph {et~al.}(2010)\citenamefont {Werner},
  \citenamefont {Oka}, \citenamefont {Eckstein},\ and\ \citenamefont
  {Millis}}]{WernerOkaEcksteinMillis:10}%
  \BibitemOpen
  \bibfield  {author} {\bibinfo {author} {\bibfnamefont {P.}~\bibnamefont
  {Werner}}, \bibinfo {author} {\bibfnamefont {T.}~\bibnamefont {Oka}},
  \bibinfo {author} {\bibfnamefont {M.}~\bibnamefont {Eckstein}}, \ and\
  \bibinfo {author} {\bibfnamefont {A.~J.}\ \bibnamefont {Millis}},\ }\href
  {\doibase 10.1103/PhysRevB.81.035108} {\bibfield  {journal} {\bibinfo
  {journal} {Phys. Rev. B}\ }\textbf {\bibinfo {volume} {81}},\ \bibinfo
  {pages} {035108} (\bibinfo {year} {2010})}\BibitemShut {NoStop}%
\bibitem [{\citenamefont {Werner}\ \emph {et~al.}(2009)\citenamefont {Werner},
  \citenamefont {Oka},\ and\ \citenamefont {Millis}}]{WernerOkaMillis:09}%
  \BibitemOpen
  \bibfield  {author} {\bibinfo {author} {\bibfnamefont {P.}~\bibnamefont
  {Werner}}, \bibinfo {author} {\bibfnamefont {T.}~\bibnamefont {Oka}}, \ and\
  \bibinfo {author} {\bibfnamefont {A.~J.}\ \bibnamefont {Millis}},\ }\href
  {\doibase 10.1103/PhysRevB.79.035320} {\bibfield  {journal} {\bibinfo
  {journal} {Phys. Rev. B}\ }\textbf {\bibinfo {volume} {79}},\ \bibinfo
  {pages} {035320} (\bibinfo {year} {2009})}\BibitemShut {NoStop}%
\bibitem [{\citenamefont {Droghetti}\ and\ \citenamefont
  {Rungger}(2017)}]{DroghettiRunger:16}%
  \BibitemOpen
  \bibfield  {author} {\bibinfo {author} {\bibfnamefont {A.}~\bibnamefont
  {Droghetti}}\ and\ \bibinfo {author} {\bibfnamefont {I.}~\bibnamefont
  {Rungger}},\ }\href {\doibase 10.1103/PhysRevB.95.085131} {\bibfield
  {journal} {\bibinfo  {journal} {Phys. Rev. B}\ }\textbf {\bibinfo {volume}
  {95}},\ \bibinfo {pages} {085131} (\bibinfo {year} {2017})}\BibitemShut
  {NoStop}%
\bibitem [{\citenamefont {Cheng}\ \emph {et~al.}(2015)\citenamefont {Cheng},
  \citenamefont {Hou}, \citenamefont {Wang}, \citenamefont {Li}, \citenamefont
  {Wei},\ and\ \citenamefont {Yan}}]{cheng2015time}%
  \BibitemOpen
  \bibfield  {author} {\bibinfo {author} {\bibfnamefont {Y.}~\bibnamefont
  {Cheng}}, \bibinfo {author} {\bibfnamefont {W.}~\bibnamefont {Hou}}, \bibinfo
  {author} {\bibfnamefont {Y.}~\bibnamefont {Wang}}, \bibinfo {author}
  {\bibfnamefont {Z.}~\bibnamefont {Li}}, \bibinfo {author} {\bibfnamefont
  {J.}~\bibnamefont {Wei}}, \ and\ \bibinfo {author} {\bibfnamefont
  {Y.}~\bibnamefont {Yan}},\ }\href
  {http://stacks.iop.org/1367-2630/17/i=3/a=033009} {\bibfield  {journal}
  {\bibinfo  {journal} {New J. Phys.}\ }\textbf {\bibinfo {volume} {17}},\
  \bibinfo {pages} {033009} (\bibinfo {year} {2015})}\BibitemShut {NoStop}%
\bibitem [{\citenamefont {Weiss}\ \emph {et~al.}(2008)\citenamefont {Weiss},
  \citenamefont {Eckel}, \citenamefont {Thorwart},\ and\ \citenamefont
  {Egger}}]{PhysRevB.77.195316}%
  \BibitemOpen
  \bibfield  {author} {\bibinfo {author} {\bibfnamefont {S.}~\bibnamefont
  {Weiss}}, \bibinfo {author} {\bibfnamefont {J.}~\bibnamefont {Eckel}},
  \bibinfo {author} {\bibfnamefont {M.}~\bibnamefont {Thorwart}}, \ and\
  \bibinfo {author} {\bibfnamefont {R.}~\bibnamefont {Egger}},\ }\href
  {\doibase 10.1103/PhysRevB.77.195316} {\bibfield  {journal} {\bibinfo
  {journal} {Phys. Rev. B}\ }\textbf {\bibinfo {volume} {77}},\ \bibinfo
  {pages} {195316} (\bibinfo {year} {2008})}\BibitemShut {NoStop}%
\bibitem [{\citenamefont {Bock}\ \emph {et~al.}(2016)\citenamefont {Bock},
  \citenamefont {Liluashvili},\ and\ \citenamefont
  {Gasenzer}}]{bock2016buildup}%
  \BibitemOpen
  \bibfield  {author} {\bibinfo {author} {\bibfnamefont {S.}~\bibnamefont
  {Bock}}, \bibinfo {author} {\bibfnamefont {A.}~\bibnamefont {Liluashvili}}, \
  and\ \bibinfo {author} {\bibfnamefont {T.}~\bibnamefont {Gasenzer}},\ }\href
  {https://link.aps.org/doi/10.1103/PhysRevB.94.045108} {\bibfield  {journal}
  {\bibinfo  {journal} {Phys. Rev. B}\ }\textbf {\bibinfo {volume} {94}},\
  \bibinfo {pages} {045108} (\bibinfo {year} {2016})}\BibitemShut {NoStop}%
\bibitem [{\citenamefont {L\'opez}\ \emph {et~al.}(2001)\citenamefont
  {L\'opez}, \citenamefont {Aguado}, \citenamefont {Platero},\ and\
  \citenamefont {Tejedor}}]{LopezAguadoPlateroTejedor:01}%
  \BibitemOpen
  \bibfield  {author} {\bibinfo {author} {\bibfnamefont {R.}~\bibnamefont
  {L\'opez}}, \bibinfo {author} {\bibfnamefont {R.}~\bibnamefont {Aguado}},
  \bibinfo {author} {\bibfnamefont {G.}~\bibnamefont {Platero}}, \ and\
  \bibinfo {author} {\bibfnamefont {C.}~\bibnamefont {Tejedor}},\ }\href
  {\doibase 10.1103/PhysRevB.64.075319} {\bibfield  {journal} {\bibinfo
  {journal} {Phys. Rev. B}\ }\textbf {\bibinfo {volume} {64}},\ \bibinfo
  {pages} {075319} (\bibinfo {year} {2001})}\BibitemShut {NoStop}%
\bibitem [{\citenamefont {L\'opez}\ \emph {et~al.}(1998)\citenamefont
  {L\'opez}, \citenamefont {Aguado}, \citenamefont {Platero},\ and\
  \citenamefont {Tejedor}}]{LopezAguadoPlateroTejedor:98}%
  \BibitemOpen
  \bibfield  {author} {\bibinfo {author} {\bibfnamefont {R.}~\bibnamefont
  {L\'opez}}, \bibinfo {author} {\bibfnamefont {R.}~\bibnamefont {Aguado}},
  \bibinfo {author} {\bibfnamefont {G.}~\bibnamefont {Platero}}, \ and\
  \bibinfo {author} {\bibfnamefont {C.}~\bibnamefont {Tejedor}},\ }\href
  {\doibase 10.1103/PhysRevLett.81.4688} {\bibfield  {journal} {\bibinfo
  {journal} {Phys. Rev. Lett.}\ }\textbf {\bibinfo {volume} {81}},\ \bibinfo
  {pages} {4688} (\bibinfo {year} {1998})}\BibitemShut {NoStop}%
\bibitem [{\citenamefont {Sch{\"o}nhammer}\ \emph {et~al.}(1995)\citenamefont
  {Sch{\"o}nhammer}, \citenamefont {Gunnarsson},\ and\ \citenamefont
  {Noack}}]{SchoenhammerGunnarssonNoack:95}%
  \BibitemOpen
  \bibfield  {author} {\bibinfo {author} {\bibfnamefont {K.}~\bibnamefont
  {Sch{\"o}nhammer}}, \bibinfo {author} {\bibfnamefont {O.}~\bibnamefont
  {Gunnarsson}}, \ and\ \bibinfo {author} {\bibfnamefont {R.~M.}\ \bibnamefont
  {Noack}},\ }\href {https://link.aps.org/doi/10.1103/PhysRevB.52.2504}
  {\bibfield  {journal} {\bibinfo  {journal} {Phys. Rev. B}\ }\textbf {\bibinfo
  {volume} {52}},\ \bibinfo {pages} {2504} (\bibinfo {year}
  {1995})}\BibitemShut {NoStop}%
\bibitem [{\citenamefont {Lima}\ \emph {et~al.}(2003)\citenamefont {Lima},
  \citenamefont {Silva}, \citenamefont {Oliveira},\ and\ \citenamefont
  {Capelle}}]{LimaSilvaOliveiraCapelle:03}%
  \BibitemOpen
  \bibfield  {author} {\bibinfo {author} {\bibfnamefont {N.~A.}\ \bibnamefont
  {Lima}}, \bibinfo {author} {\bibfnamefont {M.~F.}\ \bibnamefont {Silva}},
  \bibinfo {author} {\bibfnamefont {L.~N.}\ \bibnamefont {Oliveira}}, \ and\
  \bibinfo {author} {\bibfnamefont {K.}~\bibnamefont {Capelle}},\ }\href
  {\doibase 10.1103/PhysRevLett.90.146402} {\bibfield  {journal} {\bibinfo
  {journal} {Phys. Rev. Lett.}\ }\textbf {\bibinfo {volume} {90}},\ \bibinfo
  {pages} {146402} (\bibinfo {year} {2003})}\BibitemShut {NoStop}%
\bibitem [{\citenamefont {Malet}\ and\ \citenamefont
  {Gori-Giorgi}(2012)}]{MaletGoriGiorgi:12}%
  \BibitemOpen
  \bibfield  {author} {\bibinfo {author} {\bibfnamefont {F.}~\bibnamefont
  {Malet}}\ and\ \bibinfo {author} {\bibfnamefont {P.}~\bibnamefont
  {Gori-Giorgi}},\ }\href {\doibase 10.1103/PhysRevLett.109.246402} {\bibfield
  {journal} {\bibinfo  {journal} {Phys. Rev. Lett.}\ }\textbf {\bibinfo
  {volume} {109}},\ \bibinfo {pages} {246402} (\bibinfo {year}
  {2012})}\BibitemShut {NoStop}%
\bibitem [{\citenamefont {Lorenzana}\ \emph {et~al.}(2012)\citenamefont
  {Lorenzana}, \citenamefont {Ying},\ and\ \citenamefont
  {Brosco}}]{LorenzanaYingBrosco:12}%
  \BibitemOpen
  \bibfield  {author} {\bibinfo {author} {\bibfnamefont {J.}~\bibnamefont
  {Lorenzana}}, \bibinfo {author} {\bibfnamefont {Z.-J.}\ \bibnamefont {Ying}},
  \ and\ \bibinfo {author} {\bibfnamefont {V.}~\bibnamefont {Brosco}},\ }\href
  {\doibase 10.1103/PhysRevB.86.075131} {\bibfield  {journal} {\bibinfo
  {journal} {Phys. Rev. B}\ }\textbf {\bibinfo {volume} {86}},\ \bibinfo
  {pages} {075131} (\bibinfo {year} {2012})}\BibitemShut {NoStop}%
\bibitem [{\citenamefont {Mirtschink}\ \emph {et~al.}(2013)\citenamefont
  {Mirtschink}, \citenamefont {Seidl},\ and\ \citenamefont
  {Gori-Giorgi}}]{MirtschinkSeidlGoriGiorgi:13}%
  \BibitemOpen
  \bibfield  {author} {\bibinfo {author} {\bibfnamefont {A.}~\bibnamefont
  {Mirtschink}}, \bibinfo {author} {\bibfnamefont {M.}~\bibnamefont {Seidl}}, \
  and\ \bibinfo {author} {\bibfnamefont {P.}~\bibnamefont {Gori-Giorgi}},\
  }\href {\doibase 10.1103/PhysRevLett.111.126402} {\bibfield  {journal}
  {\bibinfo  {journal} {Phys. Rev. Lett.}\ }\textbf {\bibinfo {volume} {111}},\
  \bibinfo {pages} {126402} (\bibinfo {year} {2013})}\BibitemShut {NoStop}%
\bibitem [{\citenamefont {Verdozzi}(2008)}]{Verdozzi:08}%
  \BibitemOpen
  \bibfield  {author} {\bibinfo {author} {\bibfnamefont {C.}~\bibnamefont
  {Verdozzi}},\ }\href {\doibase 10.1103/PhysRevLett.101.166401} {\bibfield
  {journal} {\bibinfo  {journal} {Phys. Rev. Lett.}\ }\textbf {\bibinfo
  {volume} {101}},\ \bibinfo {pages} {166401} (\bibinfo {year}
  {2008})}\BibitemShut {NoStop}%
\bibitem [{\citenamefont {Kurth}\ \emph {et~al.}(2010)\citenamefont {Kurth},
  \citenamefont {Stefanucci}, \citenamefont {Khosravi}, \citenamefont
  {Verdozzi},\ and\ \citenamefont {Gross}}]{kskvg.2010}%
  \BibitemOpen
  \bibfield  {author} {\bibinfo {author} {\bibfnamefont {S.}~\bibnamefont
  {Kurth}}, \bibinfo {author} {\bibfnamefont {G.}~\bibnamefont {Stefanucci}},
  \bibinfo {author} {\bibfnamefont {E.}~\bibnamefont {Khosravi}}, \bibinfo
  {author} {\bibfnamefont {C.}~\bibnamefont {Verdozzi}}, \ and\ \bibinfo
  {author} {\bibfnamefont {E.~K.~U.}\ \bibnamefont {Gross}},\ }\href {\doibase
  10.1103/PhysRevLett.104.236801} {\bibfield  {journal} {\bibinfo  {journal}
  {Phys. Rev. Lett.}\ }\textbf {\bibinfo {volume} {104}},\ \bibinfo {pages}
  {236801} (\bibinfo {year} {2010})}\BibitemShut {NoStop}%
\bibitem [{\citenamefont {Uimonen}\ \emph {et~al.}(2011)\citenamefont
  {Uimonen}, \citenamefont {Khosravi}, \citenamefont {Stan}, \citenamefont
  {Stefanucci}, \citenamefont {Kurth}, \citenamefont {van Leeuwen},\ and\
  \citenamefont {Gross}}]{UimonenKhosraviStanStefanucciKurthLeeuwenGross:11}%
  \BibitemOpen
  \bibfield  {author} {\bibinfo {author} {\bibfnamefont {A.-M.}\ \bibnamefont
  {Uimonen}}, \bibinfo {author} {\bibfnamefont {E.}~\bibnamefont {Khosravi}},
  \bibinfo {author} {\bibfnamefont {A.}~\bibnamefont {Stan}}, \bibinfo {author}
  {\bibfnamefont {G.}~\bibnamefont {Stefanucci}}, \bibinfo {author}
  {\bibfnamefont {S.}~\bibnamefont {Kurth}}, \bibinfo {author} {\bibfnamefont
  {R.}~\bibnamefont {van Leeuwen}}, \ and\ \bibinfo {author} {\bibfnamefont
  {E.~K.~U.}\ \bibnamefont {Gross}},\ }\href
  {https://link.aps.org/doi/10.1103/PhysRevB.84.115103} {\bibfield  {journal}
  {\bibinfo  {journal} {Phys. Rev. B}\ }\textbf {\bibinfo {volume} {84}},\
  \bibinfo {pages} {115103} (\bibinfo {year} {2011})}\BibitemShut {NoStop}%
\bibitem [{\citenamefont {Kartsev}\ \emph {et~al.}(2013)\citenamefont
  {Kartsev}, \citenamefont {Karlsson}, \citenamefont {Privitera},\ and\
  \citenamefont {Verdozzi}}]{KKPV.2013}%
  \BibitemOpen
  \bibfield  {author} {\bibinfo {author} {\bibfnamefont {A.}~\bibnamefont
  {Kartsev}}, \bibinfo {author} {\bibfnamefont {D.}~\bibnamefont {Karlsson}},
  \bibinfo {author} {\bibfnamefont {A.}~\bibnamefont {Privitera}}, \ and\
  \bibinfo {author} {\bibfnamefont {C.}~\bibnamefont {Verdozzi}},\ }\href
  {\doibase doi:10.1038/srep02570} {\bibfield  {journal} {\bibinfo  {journal}
  {Sci. Rep.}\ }\textbf {\bibinfo {volume} {3}},\ \bibinfo {pages} {2570}
  (\bibinfo {year} {2013})}\BibitemShut {NoStop}%
\bibitem [{\citenamefont {Hodgson}\ \emph {et~al.}(2013)\citenamefont
  {Hodgson}, \citenamefont {Ramsden}, \citenamefont {Chapman}, \citenamefont
  {Lillystone},\ and\ \citenamefont
  {Godby}}]{HodgsonRamsdenChapmanLillystoneGodby:13}%
  \BibitemOpen
  \bibfield  {author} {\bibinfo {author} {\bibfnamefont {M.~J.~P.}\
  \bibnamefont {Hodgson}}, \bibinfo {author} {\bibfnamefont {J.~D.}\
  \bibnamefont {Ramsden}}, \bibinfo {author} {\bibfnamefont {J.~B.~J.}\
  \bibnamefont {Chapman}}, \bibinfo {author} {\bibfnamefont {P.}~\bibnamefont
  {Lillystone}}, \ and\ \bibinfo {author} {\bibfnamefont {R.~W.}\ \bibnamefont
  {Godby}},\ }\href {\doibase 10.1103/PhysRevB.88.241102} {\bibfield  {journal}
  {\bibinfo  {journal} {Phys. Rev. B}\ }\textbf {\bibinfo {volume} {88}},\
  \bibinfo {pages} {241102} (\bibinfo {year} {2013})}\BibitemShut {NoStop}%
\bibitem [{\citenamefont {Hodgson}\ \emph {et~al.}(2016)\citenamefont
  {Hodgson}, \citenamefont {Ramsden},\ and\ \citenamefont
  {Godby}}]{HodgsonRamsdenGodby:16}%
  \BibitemOpen
  \bibfield  {author} {\bibinfo {author} {\bibfnamefont {M.~J.~P.}\
  \bibnamefont {Hodgson}}, \bibinfo {author} {\bibfnamefont {J.~D.}\
  \bibnamefont {Ramsden}}, \ and\ \bibinfo {author} {\bibfnamefont {R.~W.}\
  \bibnamefont {Godby}},\ }\href {\doibase 10.1103/PhysRevB.93.155146}
  {\bibfield  {journal} {\bibinfo  {journal} {Phys. Rev. B}\ }\textbf {\bibinfo
  {volume} {93}},\ \bibinfo {pages} {155146} (\bibinfo {year}
  {2016})}\BibitemShut {NoStop}%
\bibitem [{\citenamefont {Stefanucci}\ and\ \citenamefont
  {Kurth}(2015)}]{StefanucciKurth:15}%
  \BibitemOpen
  \bibfield  {author} {\bibinfo {author} {\bibfnamefont {G.}~\bibnamefont
  {Stefanucci}}\ and\ \bibinfo {author} {\bibfnamefont {S.}~\bibnamefont
  {Kurth}},\ }\href {\doibase DOI: 10.1021/acs.nanolett.5b03294} {\bibfield
  {journal} {\bibinfo  {journal} {Nano~Lett.}\ }\textbf {\bibinfo {volume}
  {15}},\ \bibinfo {pages} {8020} (\bibinfo {year} {2015})}\BibitemShut
  {NoStop}%
\bibitem [{\citenamefont {Kurth}\ and\ \citenamefont
  {Stefanucci}(2016)}]{KurthStefanucci:16}%
  \BibitemOpen
  \bibfield  {author} {\bibinfo {author} {\bibfnamefont {S.}~\bibnamefont
  {Kurth}}\ and\ \bibinfo {author} {\bibfnamefont {G.}~\bibnamefont
  {Stefanucci}},\ }\href {\doibase 10.1103/PhysRevB.94.241103} {\bibfield
  {journal} {\bibinfo  {journal} {Phys. Rev. B}\ }\textbf {\bibinfo {volume}
  {94}},\ \bibinfo {pages} {241103(R)} (\bibinfo {year} {2016})}\BibitemShut
  {NoStop}%
\bibitem [{\citenamefont {Kurth}\ and\ \citenamefont
  {Stefanucci}(2017)}]{KurthStefanucci:17}%
  \BibitemOpen
  \bibfield  {author} {\bibinfo {author} {\bibfnamefont {S.}~\bibnamefont
  {Kurth}}\ and\ \bibinfo {author} {\bibfnamefont {G.}~\bibnamefont
  {Stefanucci}},\ }\href {http://stacks.iop.org/0953-8984/29/i=41/a=413002}
  {\bibfield  {journal} {\bibinfo  {journal} {J. Phys.: Condens. Matter}\
  }\textbf {\bibinfo {volume} {29}},\ \bibinfo {pages} {413002} (\bibinfo
  {year} {2017})}\BibitemShut {NoStop}%
\bibitem [{\citenamefont {Ullrich}(2012)}]{Ullrich:12}%
  \BibitemOpen
  \bibfield  {author} {\bibinfo {author} {\bibfnamefont {C.}~\bibnamefont
  {Ullrich}},\ }\href@noop {} {\emph {\bibinfo {title} {Time-Dependent
  Density-Functional Theory}}}\ (\bibinfo  {publisher} {Oxford University
  Press},\ \bibinfo {address} {Oxford},\ \bibinfo {year} {2012})\BibitemShut
  {NoStop}%
\bibitem [{\citenamefont {Maitra}(2016)}]{Maitra:16}%
  \BibitemOpen
  \bibfield  {author} {\bibinfo {author} {\bibfnamefont {N.~T.}\ \bibnamefont
  {Maitra}},\ }\href {\doibase http://dx.doi.org/10.1063/1.4953039} {\bibfield
  {journal} {\bibinfo  {journal} {J. Chem. Phys.}\ }\textbf {\bibinfo {volume}
  {144}},\ \bibinfo {pages} {220901} (\bibinfo {year} {2016})}\BibitemShut
  {NoStop}%
\bibitem [{\citenamefont {Kurth}\ \emph {et~al.}(2005)\citenamefont {Kurth},
  \citenamefont {Stefanucci}, \citenamefont {Almbladh}, \citenamefont {Rubio},\
  and\ \citenamefont {Gross}}]{ksarg.2005}%
  \BibitemOpen
  \bibfield  {author} {\bibinfo {author} {\bibfnamefont {S.}~\bibnamefont
  {Kurth}}, \bibinfo {author} {\bibfnamefont {G.}~\bibnamefont {Stefanucci}},
  \bibinfo {author} {\bibfnamefont {C.-O.}\ \bibnamefont {Almbladh}}, \bibinfo
  {author} {\bibfnamefont {A.}~\bibnamefont {Rubio}}, \ and\ \bibinfo {author}
  {\bibfnamefont {E.~K.~U.}\ \bibnamefont {Gross}},\ }\href
  {https://link.aps.org/doi/10.1103/PhysRevB.72.035308} {\bibfield  {journal}
  {\bibinfo  {journal} {Phys. Rev. B}\ }\textbf {\bibinfo {volume} {72}},\
  \bibinfo {pages} {035308} (\bibinfo {year} {2005})}\BibitemShut {NoStop}%
\bibitem [{\citenamefont {Dittmann}\ \emph {et~al.}()\citenamefont {Dittmann},
  \citenamefont {Splettstoesser},\ and\ \citenamefont
  {Helbig}}]{DittmannSplettstoesserHelbig}%
  \BibitemOpen
  \bibfield  {author} {\bibinfo {author} {\bibfnamefont {N.}~\bibnamefont
  {Dittmann}}, \bibinfo {author} {\bibfnamefont {J.}~\bibnamefont
  {Splettstoesser}}, \ and\ \bibinfo {author} {\bibfnamefont {N.}~\bibnamefont
  {Helbig}},\ }\href@noop {} {}\bibinfo {note} {~arXiv:cond-mat/1706.04547
  (2017)}\BibitemShut {NoStop}%
\bibitem [{\citenamefont {Jauho}\ \emph {et~al.}(1994)\citenamefont {Jauho},
  \citenamefont {Wingreen},\ and\ \citenamefont {Meir}}]{JauhoWingreenMeir:94}%
  \BibitemOpen
  \bibfield  {author} {\bibinfo {author} {\bibfnamefont {A.-P.}\ \bibnamefont
  {Jauho}}, \bibinfo {author} {\bibfnamefont {N.~S.}\ \bibnamefont {Wingreen}},
  \ and\ \bibinfo {author} {\bibfnamefont {Y.}~\bibnamefont {Meir}},\ }\href
  {\doibase 10.1103/PhysRevB.50.5528} {\bibfield  {journal} {\bibinfo
  {journal} {Phys. Rev. B}\ }\textbf {\bibinfo {volume} {50}},\ \bibinfo
  {pages} {5528} (\bibinfo {year} {1994})}\BibitemShut {NoStop}%
\bibitem [{\citenamefont {Goldin}\ and\ \citenamefont
  {Avishai}(1998)}]{GoldinAvishai:98}%
  \BibitemOpen
  \bibfield  {author} {\bibinfo {author} {\bibfnamefont {Y.}~\bibnamefont
  {Goldin}}\ and\ \bibinfo {author} {\bibfnamefont {Y.}~\bibnamefont
  {Avishai}},\ }\href {\doibase 10.1103/PhysRevLett.81.5394} {\bibfield
  {journal} {\bibinfo  {journal} {Phys. Rev. Lett.}\ }\textbf {\bibinfo
  {volume} {81}},\ \bibinfo {pages} {5394} (\bibinfo {year}
  {1998})}\BibitemShut {NoStop}%
\bibitem [{\citenamefont {Goldin}\ and\ \citenamefont
  {Avishai}(2000)}]{GoldinAvishai:00}%
  \BibitemOpen
  \bibfield  {author} {\bibinfo {author} {\bibfnamefont {Y.}~\bibnamefont
  {Goldin}}\ and\ \bibinfo {author} {\bibfnamefont {Y.}~\bibnamefont
  {Avishai}},\ }\href {\doibase 10.1103/PhysRevB.61.16750} {\bibfield
  {journal} {\bibinfo  {journal} {Phys. Rev. B}\ }\textbf {\bibinfo {volume}
  {61}},\ \bibinfo {pages} {16750} (\bibinfo {year} {2000})}\BibitemShut
  {NoStop}%
\bibitem [{\citenamefont {Elzerman}\ \emph {et~al.}(2000)\citenamefont
  {Elzerman}, \citenamefont {De~Franceschi}, \citenamefont {Goldhaber-Gordon},
  \citenamefont {van~der Wiel},\ and\ \citenamefont
  {Kouwenhoven}}]{Elzerman:00}%
  \BibitemOpen
  \bibfield  {author} {\bibinfo {author} {\bibfnamefont {J.~M.}\ \bibnamefont
  {Elzerman}}, \bibinfo {author} {\bibfnamefont {S.}~\bibnamefont
  {De~Franceschi}}, \bibinfo {author} {\bibfnamefont {D.}~\bibnamefont
  {Goldhaber-Gordon}}, \bibinfo {author} {\bibfnamefont {W.~G.}\ \bibnamefont
  {van~der Wiel}}, \ and\ \bibinfo {author} {\bibfnamefont {L.~P.}\
  \bibnamefont {Kouwenhoven}},\ }\href {\doibase 10.1023/A:1004694017738}
  {\bibfield  {journal} {\bibinfo  {journal} {J. Low Temp. Phys.}\ }\textbf
  {\bibinfo {volume} {118}},\ \bibinfo {pages} {375} (\bibinfo {year}
  {2000})}\BibitemShut {NoStop}%
\end{thebibliography}
